\documentstyle[aaspp4, 11pt]{article}

\renewcommand{\vec}[1]{\mbox{\boldmath $\displaystyle #1$}}
\newcommand{\grad}{\vec{\nabla}}

\newcommand{\vdot}{\vec{\cdot}}
\newcommand{\vcross}{\vec{\times}}
\newcommand{\divr}{\grad\vdot\,}
\newcommand{\curl}{\grad\vcross\,}

\newcommand{\avZ}{\langle Z\rangle}

\begin{document}

\title{Magnetic Screening in Accreting Neutron Stars}

\author{Andrew Cumming, Ellen Zweibel\altaffilmark{1}, and Lars Bildsten\altaffilmark{2}}
\affil{Institute for Theoretical Physics, Kohn Hall,
University of California, Santa Barbara, CA 93106; cumming@itp.ucsb.edu; zweibel@solarz.colorado.edu; bildsten@itp.ucsb.edu}

\altaffiltext{1}{Permanent address: JILA, U. Colorado, Boulder CO
80309} \altaffiltext{2}{Also: Department of Physics, University of
California, Santa Barbara, CA 93106}

\centerline{Submitted to The Astrophysical Journal, 8th February 2001}

\begin{abstract}

We investigate whether the magnetic field of an accreting neutron star
may be diamagnetically screened by the accreted matter. We assume the
freshly accumulated material is unmagnetized, and calculate the rate
at which the intrinsic stellar magnetic flux is transported into it
via Ohmic diffusion from below. For very high accretion rates $\dot M$
(larger than the Eddington rate $\dot M_{\rm Edd}$), Brown and
Bildsten have shown that the liquid ocean and outer crust of the
neutron star are built up on a timescale much shorter than the Ohmic
penetration time. We confirm this result and go further to calculate
the screening, both in this limit and at lower accretion rates, where
the screening declines. Considering the liquid ocean and outer crust,
we find that the Ohmic diffusion and accretion timescales are equal
for $\dot M\approx 0.1 \dot M_{\rm Edd}$.

We calculate the one-dimensional steady-state magnetic field profiles,
and show that the magnetic field strength decreases as one moves up
through the outer crust and ocean by $n$ orders of magnitude, where
$n\approx \dot M/0.02\ \dot M_{\rm Edd}$. We show that these profiles
are unstable to buoyancy instabilities when $B\gtrsim
10^{10}$--$10^{11}\ {\rm G}$ in the ocean. This provides a new limit
to the possible strength of any buried field. Most steadily accreting
neutron stars in low-mass X-ray binaries in our Galaxy accrete at
rates where screening would be effective if the accretion and magnetic
geometry is as we portray. However, we cannot definitively state that
the magnetic fields of these objects are screened at high levels. If
screened, then the underlying field will emerge after accretion halts,
on a timescale of only 100--1000 years, set by the Ohmic diffusion
time across the outer crust.  The definitive statement we can make is
that magnetic screening is ineffective for $\dot M<10^{-2}\dot M_{\rm
Edd}$, so that, no matter how the accreted material joins onto the
star, the underlying stellar field should always be evident. In this
respect, we point out the only known persistently-pulsing accreting
X-ray millisecond pulsar, SAX J1808.4$-$3658, has an accretion rate of
$\dot M\sim 10^{-3}\dot M_{\rm Edd}$, far below the realm where
magnetic screening can play a role.

\end{abstract}


\section{Introduction}

Accreting neutron stars fall generally into two classes (see White,
Nagase, \& Parmar 1995). Those in high mass X-ray binaries (HMXBs) are
X-ray pulsars, with dipole magnetic fields $B\sim 10^{12}\ {\rm G}$,
large enough to disrupt the accretion flow and channel it onto the
polar caps. Most of the neutron stars in low mass X-ray binaries
(LMXBs), however, show no direct evidence for a magnetic field
($B\lesssim 10^{10}\ {\rm G}$). The accretion disk is believed to
extend almost to the stellar surface, allowing these neutron stars to
be spun up to millisecond periods by accretion. Thus LMXB neutron
stars are favored candidates for the progenitors of millisecond radio
pulsars (Bhattacharya 1995), and as such are expected to have magnetic
fields $B\sim 10^8$--$10^9\ {\rm G}$.

Only one LMXB so far has shown a pulse indicative of
magnetically-controlled accretion, the newly-discovered transient
SAX~J1808.4-3658. This source shows persistent 401 Hz pulsations,
making it the first accretion-powered millisecond X-ray pulsar
(Wijnands \& van der Klis 1998; Chakrabarty \& Morgan
1998). Interpreting the pulsations as disruption of the accretion flow
by a magnetosphere leads to an estimated magnetic field strength of
$10^8$--$10^9\ {\rm G}$ (Psaltis \& Chakrabarty 1999), suggesting that
this source will become a millisecond radio pulsar once accretion
ceases (Bildsten \& Chakrabarty 2001).

The origin of such low external magnetic fields $\lesssim 10^{10}\
{\rm G}$ is not understood, but is believed to be directly connected
with accretion of matter onto the neutron star. This suggestion is
based on the observations that isolated neutron stars show no evidence
for dramatic field decay, and that low field radio pulsars occur
almost exclusively in binaries (Bhattacharya \& Srinivasan
1995). Utilizing binary evolution calculations, Taam \& van den Heuvel
(1986) studied a number of systems, finding that magnetic field
strength decreases with the amount of accreted matter, although Wijers
(1997) argues against this interpretation. A number of different
mechanisms for achieving this reduction in magnetic field strength
have been proposed (see Bhattacharya \& Srinivasan 1995 for a review).

One possibility that could explain the low dipole fields and the lack
of pulsations from these sources is that the external dipole magnetic
field is ``buried'' or ``screened'' by the accreted
matter. Bisnovatyi-Kogan \& Komberg (1974) suggested that pulsars in
binaries may have low magnetic fields because of this effect. Romani
(1990, 1995) outlined a picture in which matter initially accretes
onto the polar cap, eventually spreads under the weight of the
accumulated matter, and advects magnetic field inwards. He proposed
that the magnetic flux would be trapped in the crust and screened by
further accretion of diamagnetic matter.

In this paper, we test this possibility. We consider accretion of
unmagnetized matter onto a neutron star with a buried magnetic field,
and ask whether the field remains buried, or whether the newly
accreted material becomes magnetized faster than the rate at which it
is accreted, allowing the intrinsic stellar field to ``leak out''.  We
study the competition between downwards compression and advection by
accretion at the local rate, $\dot m$, and upwards Ohmic diffusion. We
carry out this calculation for the ocean and outer crust of the
neutron star that are in steady-state once accretion has been
occurring for an extended period. We show that, in steady-state, the
magnetic field strength decreases through the outer crust and ocean by
$n\approx \dot m/0.02\ \dot m_{\rm Edd}$ orders of magnitude, where
$\dot m_{\rm Edd}$ is the Eddington accretion rate. We show that these
profiles are unstable to buoyancy instabilities in the ocean when
$B\gtrsim 10^{10}$--$10^{11}\ {\rm G}$, giving a new limit to the
strength of any buried field.

We find that screening could be effective at the accretion rates of
most steadily accreting neutron stars. Unfortunately, because of the
simplified magnetic field and accretion geometry we adopt in this
paper, we cannot definitively state that these objects have a screened
field. However, if so, we show that after accretion ceases, we expect
the dipole field to emerge on a timescale of 100--1000 years, set by
the Ohmic diffusion time across the outer crust. It therefore seems
unlikely that screening alone can explain the low magnetic fields of
the millisecond radio pulsars.

Our robust conclusion is that it is not possible to diamagnetically
screen the field in the crust for accretion rates $\lesssim 0.02\ \dot
m_{\rm Edd}$.  The lack of screening at low accretion rates may
explain why SAX~J1808.4-3658 is the only LMXB so far to show
persistent pulsations. Consideration of both the observations and
evolutionary state of this binary (Chakrabarty \& Morgan 1998; Bildsten
\& Chakrabarty 2001) makes it clear that the time-averaged accretion
rate is unusually low, $\dot m\approx 10^{-3}\ \dot m_{\rm Edd}$, far
below the realm where screening can occur.

Previous authors have considered the competition between accretion and
Ohmic diffusion, but in different contexts. Brown \& Bildsten (1998)
compared the Ohmic diffusion and advection times in the atmosphere,
ocean and crust of the polar cap of an accreting X-ray pulsar. In this
case, the accreted matter arrives on the star already threaded by the
magnetic field, and may spread laterally if Ohmic diffusion is rapid
enough. Considering magnetic field evolution in accreting neutron
stars, Konar \& Bhattacharya (1997) assumed that the currents
supporting the magnetic field were confined to the crust, and
calculated the competing effects of Ohmic dissipation, and advection
of magnetic flux into denser regions with longer dissipation times
(and eventually into the superconducting core).

Another process which acts to transport magnetic flux outwards is
thermomagnetic drift (e.g., Urpin \& Yakovlev 1980; Geppert \& Urpin
1994). In this paper, we consider the liquid ocean and solid crust of
the neutron star, for which we find that thermomagnetic drift is
unimportant; the magnetic field profile is set by the competition
between Ohmic diffusion and accretion.  However, in the H/He layer
which overlies the ocean, we find that thermomagnetic drift dominates
the transport of magnetic field. The evolution of the magnetic field
in this layer, which periodically undergoes a thermonuclear
instability giving rise to Type I X-ray bursts, will be discussed in a
separate paper.

We start by describing the thermal structure of the ocean and crust,
and what sets the electrical conductivity there (\S 2). In \S3, we
discuss the evolution of the magnetic field under the joint action of
accretion and Ohmic diffusion, and in \S 4 evaluate the characteristic
timescales for these processes. In \S 5, we compute steady state
magnetic profiles. We show that the magnetic field decreases with
height in the ocean, by an amount that depends strongly on the local
accretion rate, $\dot m$. In \S 6, we consider the stability of the
steady-state profiles to magnetic buoyancy instabilities. We summarize
our results and conclude in \S 7 with the implications for steadily
and transiently accreting neutron stars.

\section{Microphysics and Structure of the Ocean and Crust}
\label{sec:layer}

In this section, we outline the structure of the outer layers of the
neutron star, and describe how we calculate the electrical
conductivity.

\subsection{Thermal Structure}

Most neutron stars in LMXBs accrete hydrogen and helium from their
companions at rates $\dot M\sim 10^{-11}-10^{-8} M_\odot \ {\rm
yr^{-1}}$. Since we are concerned with the outermost layers of
thickness $\lesssim 100\ {\rm m}$, small compared to the radius, we
adopt plane parallel coordinates and measure the accretion rate as a
local rate $\dot m$ (units ${\rm g\ cm^{-2}\ s^{-1}}$). The local
Eddington accretion rate is $\dot m_{\rm Edd}=2m_pc/(1+X)R\sigma_{\rm
T}$, where $\sigma_{\rm T}$ is the Thomson scattering cross-section,
$m_p$ is the proton mass, $c$ is the speed of light, $R$ is the
stellar radius, and $X$ is the hydrogen mass fraction. We use the
Eddington accretion rate for solar composition ($X=0.71$) and
$R=10$~km, $\dot m_{\rm Edd}\equiv 8.8\times 10^4\ {\rm g \ cm^{-2} \
s^{-1}}$, as our basic unit for the local accretion rate. For a 10 km
neutron star, this corresponds to a global rate $\dot M_{\rm
Edd}=1.7\times 10^{-8}\,M_\odot\,{\rm yr^{-1}}$.

The neutron star atmosphere is in hydrostatic balance as the accreted
hydrogen and helium accumulates, because the downward flow speed
($v=\dot m/\rho$) is much less than the sound speed $c_s$. The
pressure varies with height as $dP/dz=-\rho g$, where the gravitational
acceleration $g$ is constant in the thin envelope. We assume in this
paper that the magnetic field does not play a role in hydrostatic
balance, i.e. $d(B^2/8\pi)/dz\ll dP/dz$ (we check this assumption in \S
6 when we consider buoyancy instabilities). A useful variable is the
column depth $y$ (units g cm$^{-2}$), defined by $dy\equiv -\rho dz$,
giving $P=gy$. As the matter accumulates, a given fluid element moves to
greater and greater column depth. We take $g_{14}\equiv g/10^{14}\ {\rm
cm\,s^{-2}}=1.9$, the Newtonian gravitational acceleration for a
$M=1.4\,M_\odot$ and $R=10\ {\rm km}$ neutron star.

The accreted H and He burn at $y\approx 10^8$--$10^9\ {\rm g\
cm^{-2}}$ and temperature $T\approx 2$--$5\times 10^8\ {\rm K}$,
producing heavy elements which form the underlying ocean and
crust. The microphysics of these layers has been discussed by Bildsten
\& Cutler (1995) and Brown \& Bildsten (1998). The pressure in the
ocean and outer crust is mostly provided by degenerate electrons. For
the ions, we incorporate Coulomb corrections to the equation of state
using the fit of Farouki \& Hamaguchi (1993). The crystallization
point is determined by $\Gamma=(Ze)^2/k_BTa$, where $a$ is the
interion spacing given by $4\pi a^3n_i/3=1$, $Ze$ is the ionic charge,
$n_i$ is the ion number density (we assume only one species of ion is
present). For typical conditions at top of the ocean, this is
\begin{equation}
\Gamma=12\ \rho_5^{1/3}\left({2\over T_8}\right)\,\left({Z\over
30}\right)^{5/3}\, \left({2Z\over A}\right)^{1/3},
\end{equation}
where $\rho_5\equiv\rho/10^5\ {\rm g\ cm^{-3}}$, $T_8\equiv T/10^8\
{\rm K}$, and the mass of each nucleus is $Am_p$. The ions solidify when
$\Gamma=\Gamma_m\approx 173$ (Farouki \& Hamaguchi 1993), at a depth
\begin{eqnarray}
y_{cr}\approx 2\times 10^{13}\ {\rm g\ cm^{-2}}\ 
\left({T_8\over 5}\right)^4
\left({Z\over 30}\right)^{-20/3}\nonumber\\
\left({1.9\over g_{14}}\right)
\left({\Gamma_m\over 173}\right)^4,
\end{eqnarray}
where we assume fully-relativistic electrons and use the $T=0$ Fermi
energy.

We find the thermal structure of the ocean and crust by integrating
the heat equation, 
\begin{equation}\label{eq:heat}
{dT\over dy}={3\kappa F\over 4acT^3},
\end{equation}
where $F$ is the outward heat flux, and $\kappa$ is the opacity. The
heat flux is set by heat released from compression of matter by
accretion, and from pycnonuclear reactions and electron captures deep
in the crust (Brown \& Bildsten 1998). For our models, we assume a
constant heat flux $F=150\ {\rm keV}$ per nucleon. The opacity
$\kappa$ is calculated as described by Schatz et al.~(1999) and Brown
(2000) for the crust (see discussion in \S 2.2). The temperature and
pressure at the top of the ocean are set by H/He burning in the upper
atmosphere. For accretion rates $\dot m<\dot m_{\rm Edd}$, the H/He
burns unstably, and we take the temperature and column depth at the
top of the ocean to be the conditions at He ignition, as calculated by
Cumming \& Bildsten (2000). For $\dot m=\dot m_{\rm Edd}$, we use the
steady-state burning model of Schatz et al.~(1999), giving $y=10^8\
{\rm g\ cm^{-2}}$ and $T=5\times 10^8\ {\rm K}$ at the top of the
ocean. We take the composition of the ocean and crust to be nuclei
with $Z=30$ and $A=60$. We arbitrarily stop our integrations at
$y=10^{14}\ {\rm g\ cm^{-2}}$, thus we include only the outer crust,
above neutron drip.

The upper panel of Figure 1 shows the temperature profiles for $\dot
m=0.01,0.1$ and $1 \ \dot m_{\rm Edd}$. For the $0.01$ and $0.1\ \dot
m_{\rm Edd}$ models, we mark the liquid-solid interface with a dot
(the $\dot m_{\rm Edd}$ model is still liquid at $y=10^{14}\ {\rm g\
cm^{-2}}$). In the crust, we show two models for each accretion rate,
each with a different choice for the thermal conductivity: phonon
scattering or electron-ion scattering (see \S 2.2 for discussion of
conductivity in the crust).

\subsection{Electrical Conductivity}

\begin{figure}
\epsscale{1.0}\plotone{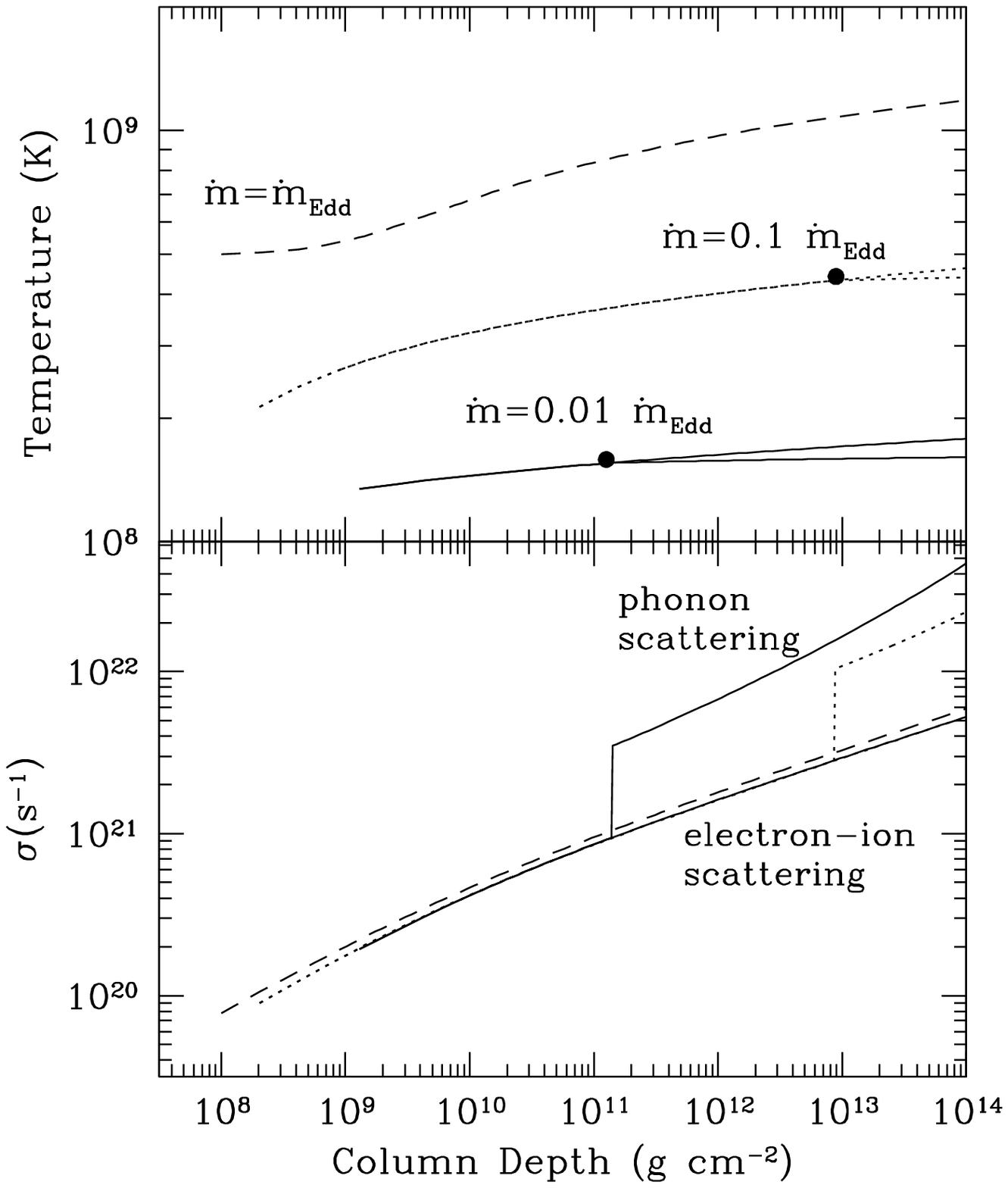}\epsscale{1.0}
\caption{Temperature and conductivity profiles in the ocean and outer
crust. We take the temperature and column depth at the top of the
ocean from the X-ray burst ignition models of Cumming \& Bildsten
(2000) for $\dot m=0.01$ (solid curves) and $0.1\ \dot m_{\rm Edd}$
(dotted curves), and from the steadily burning model of Schatz et
al.~(1999) for $\dot m=\dot m_{\rm Edd}$ (dashed curves). We assume a
single species of nuclei with $A=60$ and $Z=30$, and assume a constant
heat flux of $F=150\ {\rm keV}$ per nucleon. We indicate the
ocean-crust boundary with a filled circle (the $\dot m_{\rm Edd}$
model remains liquid for $y<10^{14}\ {\rm g\ cm^{-2}}$). For $0.01$
and $0.1\ \dot m_{\rm Edd}$, the two different curves beyond the
filled circle are for either electron-ion scattering or phonon
scattering in the crust.
\label{fig:magprof}} 
\end{figure}

In the relaxation time approximation, the electrical conductivity is
$\sigma=n_ee^2/m_\star\nu_c$, or
\begin{eqnarray}\label{eq:cond}
\sigma&=&7.6\times 10^{20}\ {\rm s^{-1}}\
\rho_5\ \left({2\over\mu_e}\right) \left({10^{16}\ {\rm s^{-1}}\over
\nu_c}\right) \left({m_e\over m_\star}\right)\nonumber
\end{eqnarray}
(e.g., Ziman 1964; Yakovlev \& Urpin 1980), where $n_e$
is the electron number density, $m_\star$ is the effective electron
mass, $\mu_e$ is the mean molecular weight per electron, and $\nu_c$
is the electron collision frequency. The lower panel of Figure
\ref{fig:magprof} shows the conductivity as a function of depth.

In the liquid ocean, electron-ion collisions set the electrical
conductivity. We use the results of Yakovlev \& Urpin (1980), who give
\begin{equation}\label{eq:YUsigma}
\sigma={8.5\times 10^{21}\ {\rm s^{-1}}\over Z\Lambda_{ei}} {x^3\over
1+x^2},
\end{equation}
where $\Lambda_{ei}\approx 1$ is the Coulomb logarithm, and $x\equiv
p_F/m_ec$ measures the electron Fermi momentum $p_F$.

The conductivity in the solid crust depends on the level of
impurity. Itoh \& Kohyama (1993) give the electron-impurity collision
frequency as
\begin{equation}\label{eq:nueQ}
\nu_{eQ}=1.75\times 10^{16}\ {\rm s^{-1}}\
{Q\over\avZ}\Lambda_{eQ}(1+x^2)^{1/2},
\end{equation}
where $Q=\sum Y_i(Z_i-\avZ)^2/\sum Y_i$ is the impurity parameter,
$\avZ=\sum Y_iZ_i/\sum Y_i=Y_e/\sum Y_i$ is the average charge (the
number abundance of species $i$ is $Y_i=X_i/A_i$, where $X_i$ is the
mass fraction), and the Coulomb logarithm is order unity. The other
contribution to the conductivity is scattering with phonons, for which
we adopt the results of Baiko \& Yakovlev (1995). For temperatures
much greater than the Debye temperature ($T\gg\Theta_D=2.4\times 10^6\
{\rm K}\ x^{3/2}(2Z/A)$), they find
\begin{equation}
\nu_{ep}= {\alpha k_BT\over\hbar}\,13\Lambda_{ep}= 1.24\times 10^{18}\
{\rm s^{-1}}\ T_8\Lambda_{ep},
\end{equation}
where $\alpha=1/137$ is the fine structure constant, and
$\Lambda_{ep}$ is of order unity. The ratio of the impurity and phonon
scattering frequencies is
\begin{equation}
{\nu_{eQ}\over\nu_{ep}}\approx 10^{-2}\ Q\ \left({3\over T_8}\right)
\left({x\over 60}\right) \left({30\over\avZ}\right)
\left({\Lambda_{eQ}\over\Lambda_{ep}}\right),
\end{equation}
where we choose typical conditions in the crust.

Previous studies have assumed a low level of impurity, $Q\ll 1$,
as might be expected for a primordial crust. In this case phonon
scattering dominates the conductivity. However, recent calculations of
the products of H/He burning in the atmosphere show that the impurity
level is probably much greater. Schatz et al.~(1999) calculated the
ashes of steady-state H/He burning at local accretion rates $\dot
m\gtrsim\dot m_{\rm Edd}$, and found $\avZ=24$ and $Q\approx 100$. At
such high impurity factors, it is not clear whether equation
(\ref{eq:nueQ}) is correct. Brown (2000) suggests that when $Q\approx
\avZ^2$, the appropriate conductivity is that for electron-ion
scattering in the liquid.

In this paper, we show results for two limiting cases for the crust: a
pure crust with only phonon scattering, and an impure crust with
electron-ion scattering calculated as for the liquid. The conductivity
in each case is shown in the lower panel of Figure 1. For each
accretion rate, we plot the two solutions for the crust: for
electron-ion scattering, the conductivity is continuous across the
liquid-solid boundary; for phonon scattering, the conductivity changes
discontinuously.\footnote{Recent conductivity calculations of Potekhin
et al.~(1999) show that the conductivity is continuous across the
melting point even in this case; however, the older results we adopt
in this paper are sufficient for our purposes.} The difference in
conductivity between these two cases is roughly a factor of 5. In \S
4, we show that the impurity level has an important effect on magnetic
field evolution in the outer crust.

\section{Evolution of the Magnetic Field} \label{sec:EVOL}

We consider a simplified geometry for the magnetic field, a plane
parallel model in which the magnetic field $\vec B$ depends only on
depth. In this case, the vertical component of the magnetic field must
be constant since $\divr\vec{B}=0$, and we set it equal to zero since
we assume the accreted matter is unmagnetized. This leaves us with a
purely horizontal field, $\vec{B}(z)=B(z)\hat{e}_x$ (we align our
$x$-axis with the horizontal component of the field).

The magnetic field evolves according to Faraday's law
\begin{equation}\label{faraday}
\frac{\partial\vec{B}}{\partial t} = -c\curl\vec{E},
\end{equation}
where the electric field $\vec{E}$ is given by Ohm's law,
\begin{equation}\label{eq:ohmslaw}
\vec{E}=-\frac{\vec{v}\times\vec{B}}{c} + \frac{\vec{J}}{\sigma}.
\end{equation}
The electric current is
\begin{equation}
\vec{J}={c\over 4\pi}\curl\vec{B}=\hat{e}_y{c\over 4\pi} {\partial
B\over \partial z},
\end{equation}
for a given profile $\vec{B}(z)$, and the downwards flow velocity is
given by continuity as $\vec{v}=-v(z)\hat{e}_z=-(\dot
m/\rho)\hat{e}_z$ (we define $v$ so that it is a positive quantity for
the downwards flow). Thus $\vec{E}=E_y\hat{e}_y$, where
\begin{equation}
E_y={vB\over c}+{c\over 4\pi\sigma}{\partial B\over \partial z},
\end{equation}
and equation (\ref{faraday}) reduces to its $\hat x$ component
\begin{equation}\label{eq:induction}
{\partial B\over\partial t}={\partial\over\partial z}
\left[vB + \eta\frac{\partial B}{\partial z}\right],
\end{equation}
where $\eta=c^2/4\pi\sigma$ is the magnetic diffusivity.

It is instructive to rewrite equation (\ref{eq:induction}) as 
\begin{equation}\label{eq:advect}
\rho{D\over Dt}\left({B\over\rho}\right) ={\partial\over\partial
z}\left(\eta{\partial B\over\partial z}\right),
\end{equation}
where $D/Dt=\partial/\partial t+\vec{v}\vdot\grad$ is the advective
derivative and we have used the continuity equation
$D\rho/Dt=-\rho\divr\vec{v}$. Equation (\ref{eq:advect}) clearly
illustrates the different processes acting to change the magnitude of
$B$. The left hand side describes advection and compression by the
downwards flow; the right hand side describes Ohmic diffusion. Without
diffusion, $B$ in a given fluid element would grow $\propto\rho$ as it
was compressed.

In this paper, we calculate only steady state solutions of
equation (\ref{eq:induction}). Since we assume that the incoming material
is unmagnetized, we set $E_y = 0$, and equation (\ref{eq:induction})
reduces to
\begin{equation}\label{induction}
\eta\frac{dB}{dz} = -vB.
\end{equation}
We rewrite this so that pressure is the independent
variable; this makes the competition between advection and diffusion
more transparent. Using hydrostatic balance, and introducing the
pressure scale height $H\equiv y/\rho= -(d\ln P/dz)^{-1}$, equation
(\ref{induction}) becomes
\begin{equation}\label{ss}
\frac{d\ln B}{d\ln P}=\left(\frac{H^2}{\eta}\right)\left(\frac{\dot
m}{y}\right) = \frac{t_{\rm diff}}{t_{\rm
accr}},
\end{equation}
where we have defined 
\begin{equation}\label{tdiff}
t_{\rm diff}\equiv\frac{H^2}{\eta},
\end{equation}
the characteristic Ohmic diffusion time across a scale height, and
\begin{equation}\label{taccr}
t_{\rm accr}\equiv\frac{y}{\dot m},
\end{equation}
the characteristic accretion flow time across a pressure scale
height. Equation (\ref{ss}) shows that, in steady-state, $t_{\rm
diff}/t_{\rm accr}$ measures the ratio of pressure scale height to
magnetic scale height. For example, if $t_{\rm diff}>t_{\rm accr}$, a
small magnetic scale height is required for Ohmic diffusion to balance
accretion.

The magnetic profile is given by integrating equation (\ref{ss}),
since all the quantities on the right hand side are known functions of
pressure (or height). We find
\begin{equation}\label{sssol}
B(P)=B(P_0)S,
\end{equation}
where $P_0$ is a reference pressure, and
\begin{equation}\label{S}
S\equiv \exp\left[-\int_{P}^{P_0}d\ln
P\left(\frac{t_{\rm diff}}{t_{\rm
accr}}\right)\right]
\end{equation}
is the amount by which the magnetic field changes due to screening
from the overlying matter. We shall refer to the quantity $S$ as the
magnetic screening factor.

The Ohmic diffusion time is independent of $B$ (see Goldreich \&
Reisenegger 1992 for an interesting discussion). Thus equation
(\ref{ss}) is linear in $B$, giving the profile of the magnetic field
independent of its overall amplitude. In the field burial picture, the
amplitude of $B$ in the crust is fixed by conserving the magnetic flux
in the original dipole field. We do not calculate the details of the
burial process in this paper; thus we treat the overall normalization
of the magnetic field in the crust as a free parameter.

For our assumed field geometry, the Hall term in Ohm's law exactly
vanishes. More generally, the magnitude of the Hall term depends on
the ratio of the electron cyclotron frequency, $\omega_e=Be/m_\star
c$, to the electron collision frequency, $\nu_c=\tau^{-1}$. For a
typical value of $\nu_c$, this is
\begin{equation}\label{omegatau}
\omega_e\tau=\left({B\over 5.6\times 10^8\ {\rm G}}\right)
\left({10^{16}\ {\rm s^{-1}}\over\nu_c}\right) \left({m_e\over
m_\star}\right),
\end{equation}
where $m_\star=E_e/c^2$ is the effective mass of the electron.
Typically, $\omega_e\tau>1$ requires $B\gtrsim 10^{11}\ {\rm G}$ in
the ocean and crust; we show in \S \ref{sec:stability} that such large
fields are buoyantly unstable. In addition, we neglect terms in Ohm's
law due to thermomagnetic effects (e.g., Urpin \& Yakovlev 1980);
these are unimportant in the ocean and outer crust.

\section{Accretion vs. Ohmic Diffusion}
\label{sec:DIFF}

\begin{figure}
\epsscale{1.0}\plotone{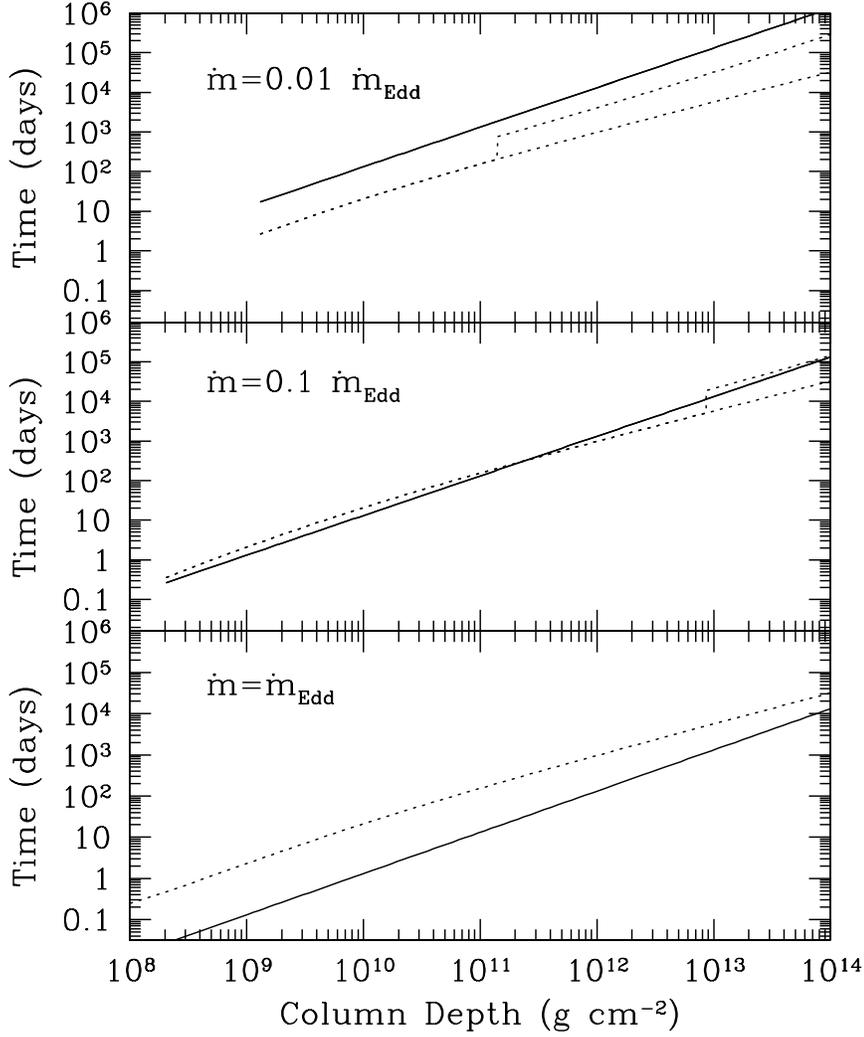}\epsscale{1.0}
\caption{ Accretion time (solid line) and Ohmic diffusion time (dotted
line) across a scale height, for (top to bottom panels) $\dot m=0.01$,
$0.1$ and $1\ \dot m_{\rm Edd}$. The two solutions in the
crust are for phonon or electron-ion scattering (see
Figure 1).\label{fig:magtimes}}
\end{figure}

In \S \ref{sec:EVOL}, we showed that the steady-state magnetic profile
depends directly on the ratio of accretion and Ohmic diffusion
timescales. In this section, we evaluate this ratio, before presenting
the steady-state profiles in \S \ref{sec:steady}. Figure
\ref{fig:magtimes} shows the accretion and diffusion timescales as a
function of depth, and Figure \ref{fig:ratio} shows $t_{\rm
diff}/t_{\rm accr}$. We find that the diffusion time is longer than
accretion time for $\dot m\gtrsim 0.1\ \dot m_{\rm Edd}$. We now use
simple analytic estimates to understand these results, discussing the
ocean (\S \ref{sec:DIFF1}) and crust (\S \ref{sec:DIFF3}) in turn.

\subsection{Ocean}\label{sec:DIFF1}

First, we evaluate $t_{\rm diff}=4\pi\sigma H^2/c^2$. The conductivity
is given by equation (\ref{eq:YUsigma}). The degenerate electron
pressure is $P_e=(m_ec^2)^4f(x)/24\pi^2(\hbar c)^3$ (Clayton 1983),
or, in terms of column depth $y=P/g$,
\begin{equation}\label{eq:y8degen}
y_8=3.2\ f(x) \left({1.9\over g_{14}}\right)
\end{equation}
where
\begin{equation}\label{eq:f}
f(x)=x(2x^2-3)(1+x^2)^{1/2}+3\sinh^{-1}x
\end{equation}
(Clayton 1983), and $y_8\equiv y/10^8\ {\rm g\ cm^{-2}}$. We write the
density as $\rho=\mu_en_em_p$, giving $\rho_5=9.7\ \mu_e\ x^3$. The
scale height $H=y/\rho$ is thus
\begin{equation}
H=170\ {\rm cm}\ {f(x)\over x^3} \left({1.9\over g_{14}}\right)
\left({2\over \mu_e}\right),
\end{equation}
giving
\begin{equation}\label{eq:tdiffdegen}
t_{\rm diff}=
10^5\ {\rm s}\ {f^2(x)\over x^3(1+x^2)}
\Lambda_{ei}^{-1}\left({30\over Z}\right) \left({1.9\over
g_{14}}\right)^2 \left({2\over \mu_e}\right)^2.
\end{equation}
The accretion timescale is
\begin{equation}\label{eq:taccr} 
t_{\rm accr}=
10^4\,{\rm s}\ y_8\left({\dot m\over 0.1\ \dot
m_{\rm Edd}}\right)^{-1},
\end{equation}
giving
\begin{eqnarray}\label{eq:degenratio}
{t_{\rm diff}\over t_{\rm accr}}=
3\ g(x)\,\Lambda_{ei}^{-1}\,  
\left({\dot m\over 0.1\ \dot m_{\rm Edd}}\right)
\left({2\over \mu_e}\right)^2
\nonumber\\
\left({30\over Z}\right)
\left({1.9\over g_{14}}\right),
\end{eqnarray}
where $g(x)\equiv f(x)/x^3(1+x^2)$. Thus, $t_{\rm diff}\approx t_{\rm
accr}$ for $\dot m\approx 0.1\ \dot m_{\rm Edd}$, in agreement with
Figure \ref{fig:ratio}.

In the non-relativistic and relativistic limits, the functions $f(x)$
and $g(x)$ take the limiting values
\begin{equation}
f(x)=\cases{8x^5/5 & $x\ll 1$\cr 2x^4 & $x\gg 1$ \cr }
\end{equation}
and
\begin{equation}
g(x)=\cases{8x^2/5 & $x\ll 1$\cr 2/x & $x\gg 1$ \cr }.
\end{equation}
Thus 
\begin{eqnarray}\label{eq:RAT1} \left({t_{\rm diff}\over
t_{\rm accr}}\right)_{NR}\approx
3\ y_8^{2/5}\,\Lambda_{ei}^{-1}\, \left({\dot m\over 0.1\ \dot m_{\rm
Edd}}\right) \left({2\over \mu_e}\right)^2
\nonumber\\
\left({30\over Z}\right)
\left({1.9\over g_{14}}\right)^{3/5},
\end{eqnarray}
for $x\ll 1$, and
\begin{eqnarray}\label{eq:RAT2}
\left({t_{\rm diff}\over t_{\rm accr}}\right)_{R}\approx
\ y_{12}^{-1/4}\,\Lambda_{ei}^{-1}\,
\left({\dot m\over
0.1\ \dot m_{\rm Edd}}\right) \left({2\over \mu_e}\right)^2
\nonumber\\
\left({30\over Z}\right) \left({1.9\over g_{14}}\right)^{5/4}
\end{eqnarray}
for $x\gg 1$. These analytic estimates roughly agree with our
numerical results, and explain the scaling of $t_{\rm diff}/t_{\rm
accr}$ with depth; for small $x$, $t_{\rm diff}/t_{\rm accr}$
increases with depth, whereas for large $x$, $t_{\rm diff}/t_{\rm
accr}$ decreases with depth.  The total screening factor $S$ through
the ocean is thus insensitive to the upper and lower boundaries. The
maximum value of $t_{\rm diff}/t_{\rm accr}$ occurs for intermediate
values of $y$, for which the electrons are partially relativistic
($x\sim 1$, $\rho\approx 10^6$--$10^7\ {\rm g\ cm^{-3}}$).

\subsection{Crust}\label{sec:DIFF3}

\begin{figure}
\plotone{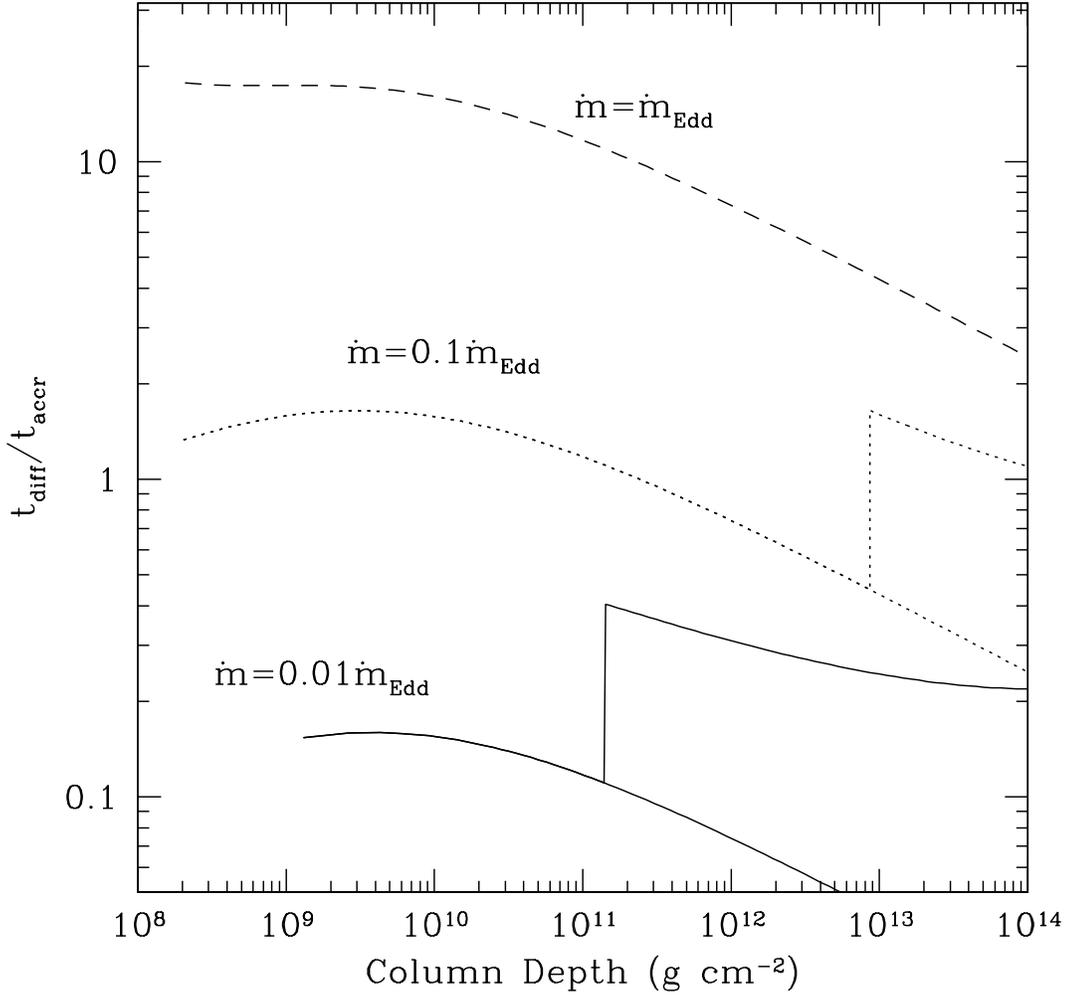}
\caption{ The ratio of Ohmic diffusion time across a pressure scale
height $t_{\rm diff}$ to the accretion flow time across a pressure
scale height $t_{\rm accr}$, for $\dot m=0.01$ (solid), $0.1$ (dotted)
and $1\ \dot m_{\rm Edd}$ (dashed line). The two solutions in the
crust are for phonon or electron-ion scattering. 
\label{fig:ratio}}
\end{figure}

We now turn to the solid crust. The behavior of $t_{\rm diff}/t_{\rm
accr}$ is quite different in the crust, depending on whether we choose
electron-ion scattering or phonon scattering. For electron-ion
scattering, $t_{\rm diff}/t_{\rm accr}\propto g(x)\propto 1/x$,
decreasing with depth, as shown in Figure \ref{fig:ratio}. For phonon
scattering,
\begin{equation}
t_{\rm diff}=2
\times 10^5\,{\rm s}\ {x^4\over T_8}\,
\left({13f_{ep}\over\Lambda_{ep}}\right)
\left({1.9\over g_{14}}\right)^2
\left({2\over\mu_e}\right)^2,
\end{equation}
giving
\begin{eqnarray}\label{eq:CR2}
\left({t_{\rm diff}\over t_{\rm accr}}\right)_{p}={3
\over T_8}\ 
\left({\dot m\over 0.1\ \dot m_{\rm Edd}}\right)
\nonumber\\
\left({1.9\over g_{14}}\right)\left({2\over\mu_e}\right)^2
\left({13f_{ep}\over\Lambda_{ep}}\right).
\end{eqnarray}
Thus $t_{\rm diff}/t_{\rm accr}$ is constant with depth in this case,
in agreement with Figure \ref{fig:ratio}. We see that, as suggested by
Schatz et al.~(1999), $t_{\rm diff}$ is much smaller in an impure
crust (electron-ion scattering) than a primordial crust (phonon
scattering).

We do not calculate the conductivity of the inner crust, below neutron
drip at $y\sim 10^{15}$--$10^{16}\ {\rm g\ cm^{-2}}$. At these depths,
Brown \& Bildsten (1998) found that $t_{\rm diff}<t_{\rm accr}$ for
$\dot m<5\dot m_{\rm Edd}$ (see their Figure 9 for a continuation of
our Figure 3 into the inner crust). Thus the inner crust does not give
a significant contribution to the screening factor.

\section{Steady-State Profiles}\label{sec:steady}

In section \ref{sec:DIFF}, we compared the timescale for Ohmic
diffusion with that for accretion. We now use the results of \S 3 to
compute the steady-state magnetic profiles, by integrating equation
(\ref{ss}).

\begin{figure}
\epsscale{1.0}\plotone{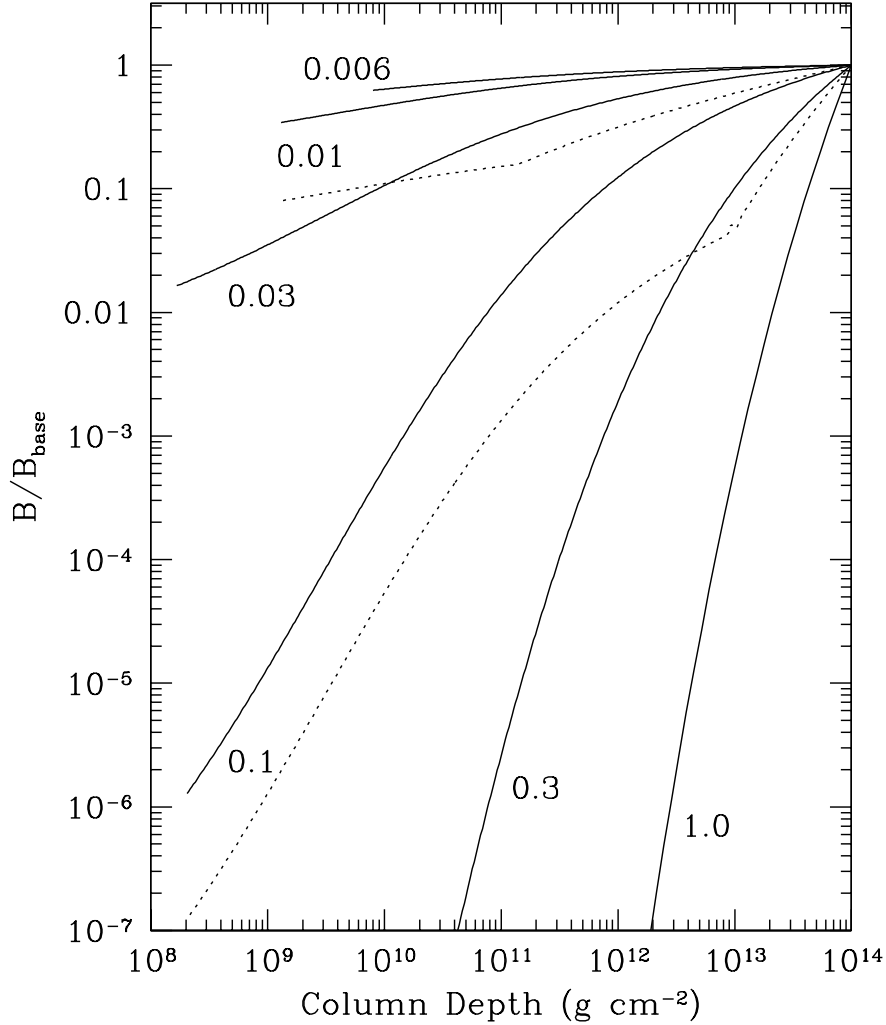}\epsscale{1.0}
\caption{Steady-state profiles of the magnetic field, for (left to
right) $\dot m=0.006, 0.01, 0.03, 0.1, 0.3$ and $1\ \dot m_{\rm
Edd}$. We plot the magnetic field relative to the value at $y=10^{14}\
{\rm g\ cm^{-2}}$. The solid curves show the profiles for electron-ion
scattering in the crust. For $\dot m=0.01$ and $0.1\ \dot m_{\rm
Edd}$, the dotted curves show the profiles for phonon scattering.
\label{fig:bfield}}
\end{figure}

Figure \ref{fig:bfield} shows the result for a range of accretion
rates. The solid lines are for models with electron-ion scattering in
the crust; for $0.01$ and $0.1 \dot m_{\rm Edd}$, the dotted lines
show models with phonon scattering. Rather than fix the overall
magnitude of the magnetic field, we plot the ratio of the magnetic
field at each depth to the magnetic field at the base ($y=10^{14}\
{\rm g\ cm^{-2}}$). As expected from inspection of equation
(\ref{ss}), we find that the magnetic and pressure scale heights are
equal for $\dot m\approx 0.1\ \dot m_{\rm Edd}$, the accretion rate at
which $t_{\rm diff}\sim t_{\rm accr}$ (\S \ref{sec:DIFF1}).

An analytic estimate of the screening factor for the ocean follows
straightforwardly from the results of \S \ref{sec:DIFF}. The ratio
$t_{\rm diff}/t_{\rm accr}$ is given by equation
(\ref{eq:degenratio}). We then substitute $d\ln P=(d\ln f/dx)\,dx$ in
equation (\ref{S}), where $f(x)$ is given by equation (\ref{eq:f}) and
$df/dx=8x^4/(1+x^2)^{1/2}$. As we discussed in \S \ref{sec:DIFF1},
most of the contribution to $S$ comes from $x\sim 1$. Integrating, we
find
\begin{eqnarray}\label{eq:Qdegen}
\ln S=2.4\ \left({\dot m\over 0.01\ \dot m_{\rm Edd}}\right)
\left({30\over Z}\right)\left({1.9\over g_{14}}\right)
\nonumber\\\left[{1\over
(1+x_t^2)^{1/2}}-{1\over (1+x_b^2)^{1/2}}\right]
\end{eqnarray}
where $x_t$ ($x_b$) is the value of $x$ at the top (bottom), and we
take $\mu_e=2$ and $\Lambda_{ei}=1$. Note that equation
(\ref{eq:Qdegen}) gives $\ln S$; the factor by which the magnetic
field changes through the atmosphere is the exponential of this value.

Using the scalings of equation (\ref{eq:Qdegen}), and adopting the
prefactor from our numerical results, we find that the magnetic field
decreases by $n$ orders of magnitude through the ocean, where
\begin{equation}
n\approx {\dot m\over 0.02\ \dot m_{\rm Edd}}\ \left({Z\over 30}\right)
\left({g_{14}\over 1.9}\right).
\end{equation}
The major contribution to $n$ comes from the ocean. Thus the
uncertainty in the conductivity of the crust (electron-ion or phonon
scattering) changes $n$ by only a small amount, as shown by comparing
the solid and dotted lines in Figure \ref{fig:bfield}.

\section{Buoyancy Instability}
\label{sec:stability}

We now investigate the stability of the steady-state magnetic profiles
to buoyancy instabilities. We first consider interchange and Parker
instabilities in \S \ref{sec:stab1}, before including the effects of
thermal diffusion in \S \ref{sec:stab2}.

\subsection{Interchange and Parker Instabilities}\label{sec:stab1}

The simplest case to consider is the interchange instability in which
a magnetic field line and associated fluid is lifted vertically,
maintaining pressure balance with its surroundings. If the new density
is less than that of the surrounding fluid, it is buoyantly
unstable. Newcomb (1961) considered the stability of a horizontal
magnetic field in a stratified atmosphere, pointing out that it is
also important to consider Parker-type modes, in which the magnetic
field lifts up, but the fluid flows back down the field lines. In the
limit of long wavelength along the field (minimizing the energy used
to bend the magnetic field lines), Newcomb showed that the Parker-type
modes may be unstable even when the interchange modes are not,
although with smaller growth rates.

Newcomb (1961) showed that the criterion for instability is
\begin{equation}
{\mathcal A}\equiv {d\ln\rho\over dz}+{\rho g\over\Gamma_1 P_g}>0,
\end{equation}
where $P_g$ is the gas pressure. Using the equation of hydrostatic
balance, $dP/dz=d(P_g+B^2/8\pi)/dz=-\rho g$, to substitute for $-\rho g$,
we find instability if (see also Acheson 1979)
\begin{equation}\label{eq:acheson}
\left({2\over\Gamma_1}\right)\left({B^2\over 8\pi P_g}\right)
\left(-{d\ln B\over dz}\right)>{N^2/g},
\end{equation}
where 
\begin{equation}\label{eq:N2}
{N^2\over g}={d\ln\rho\over dz}-{1\over\Gamma_1}{d\ln P_g\over dz}
\end{equation}
defines the Brunt V\"ais\"al\"a frequency $N^2$ (e.g., Hansen \&
Kawaler 1994). Both thermal buoyancy and composition gradients may
contribute to $N^2$ (e.g., Bildsten \& Cumming 1998). We assume a
uniform composition in the ocean, so that $N^2$ has only a thermal
piece,
\begin{equation}\label{eq:N21}
N^2={g\over H}{\chi_T\over\chi_\rho}
\left(\nabla_{\rm ad}-{d\ln T\over d\ln y}\right),
\end{equation}
where $\chi_Q\equiv\partial\ln P/\partial \ln Q$ with the other
independent thermodynamic variables held constant.

We see from equation (\ref{eq:acheson}) that the instability criterion
depends on both how quickly the magnetic field decreases with height,
and the ratio of magnetic to gas pressure, $B^2/8\pi P_g$. To
understand this, first consider a magnetic field whose strength
changes discontinuously with height. If the magnetic field above the
discontinuity is smaller than that below, pressure continuity implies
heavy fluid overlying light; a Rayleigh Taylor unstable density jump
no matter how small $B^2/8\pi P_g$. On the other hand, if magnetic
pressure dominates gas pressure, even a small magnetic field gradient
may create an unstable density profile.

\begin{figure}
\epsscale{1.0}\plotone{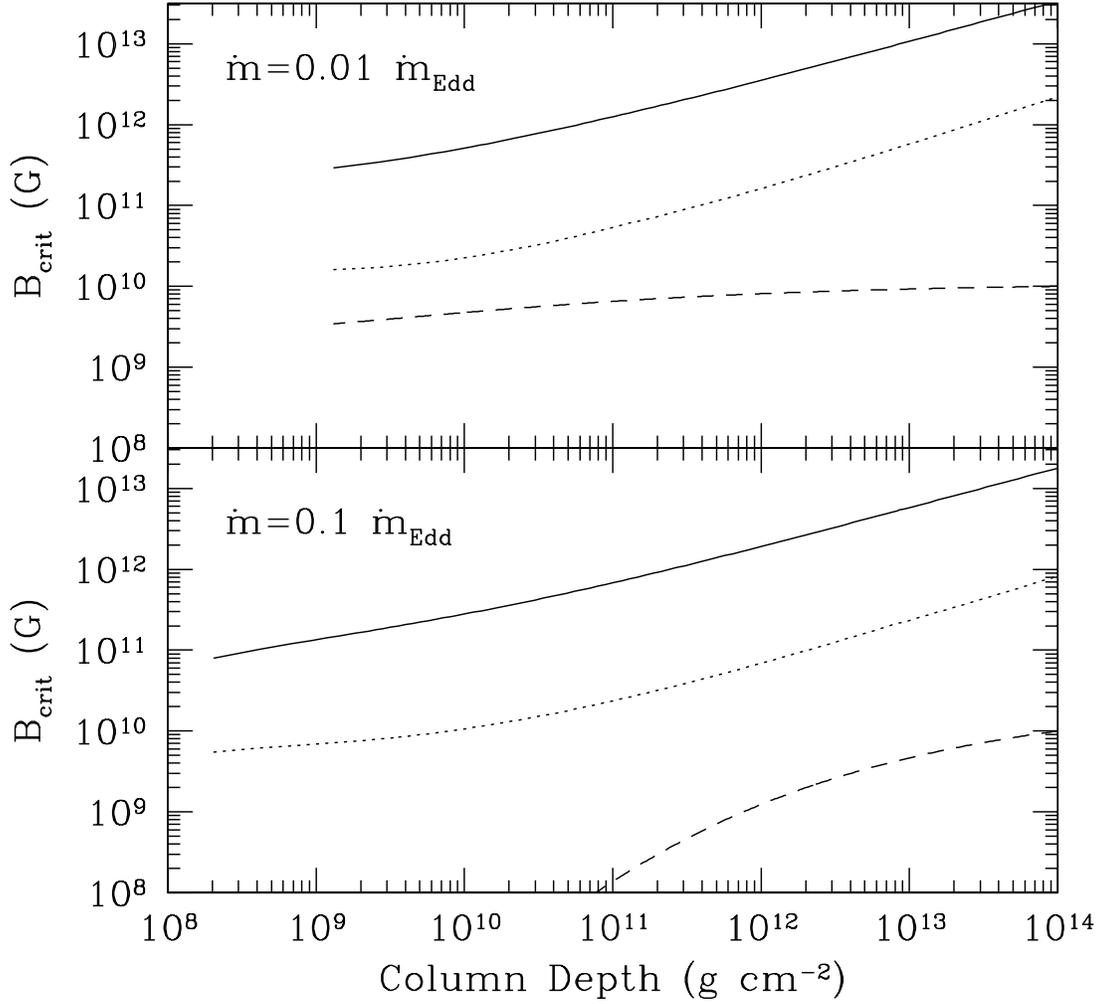}\epsscale{1.0}
\caption{ The critical magnetic field for buoyancy instability. Taking
the gradient of the steady-state magnetic profile (shown as a dashed
line, arbitrarily scaled to $10^{10}\ {\rm G}$ at the base), we
calculate the critical magnetic field at which buoyancy instability
will occur. The solid line shows the critical field ignoring thermal
diffusion, the dotted line includes thermal diffusion, which reduces
the effective thermal buoyancy. For clarity, we adopt only models with
electron-ion scattering in the crust for this figure.
\label{fig:Bcrit1}}
\end{figure}

For the steady-state magnetic profile, the gradient $d\ln B/dz$ is
given by $d\ln B/d\ln y=t_{\rm diff}/t_{\rm accr}$
(eq.~[\ref{ss}]). Inserting this gradient into equation
(\ref{eq:acheson}), we find the critical magnetic field strength $B_c$
required for instability of the steady-state profile is
\begin{equation}\label{eq:Bc}
B_c^2=4\pi\Gamma_1P_g\ \left({t_{\rm accr}\over t_{\rm diff}}\right)
\left({N^2H\over g}\right).
\end{equation}
We plot $B_c$ as a function of depth in Figure \ref{fig:Bcrit1} (solid
line). We assume $P\approx P_g$ in making this plot. The dashed line
in Figure \ref{fig:Bcrit1} shows the steady-state magnetic profile,
arbitrarily scaled to $B=10^{10}\ {\rm G}$ at the base. For clarity,
we show only models with electron-ion scattering in the crust.

In the degenerate ocean, Bildsten \& Cumming (1998) estimated the
thermal buoyancy as
\begin{equation}
{N^2H^2}\approx {3k_BT\over 8\mu_im_p},
\end{equation}
allowing us to make an analytic estimate of $B_c$. We use equation
(\ref{eq:degenratio}) for $t_{\rm diff}/t_{\rm accr}$, which gives
\begin{eqnarray}
B_c=7\times 10^{10}\ {\rm G}\ T_8^{1/2}\ 
\left[{x^3\over g(x)}\right]^{1/2}
\left({\dot m\over 0.01\ \dot m_{\rm Edd}}\right)^{-1/2},
\end{eqnarray}
where we take $\Lambda_{ei}=1$, $g_{14}=1.9$, $\Gamma_1=4/3$ and
$\mu_e=2$. For $x\ll 1$, we find $B_c$ is weakly dependent on depth,
$B_c\propto x^{1/2}\propto y^{1/10}$. For $x\gg 1$, $B_c\propto
x^{2}$, giving
\begin{equation}
B_c=4\times 10^{11}\ {\rm G}\ y_{10}^{1/2}\left({T_8\over
3}\right)^{1/2}\left({\dot m\over 0.01\ \dot m_{\rm
Edd}}\right)^{-1/2}
\end{equation}
in good agreement with our numerical results.

\subsection{Doubly-Diffusive Instability}\label{sec:stab2}

On small enough wavelengths, thermal diffusion acts to reduce thermal
gradients during the perturbation, allowing instability even for
$B<B_c$. If the ratio of magnetic to thermal diffusivities,
$\eta/{\mathcal K}$, is small, the fluid elements retain their
destabilizing magnetic buoyancy, but the stabilizing thermal buoyancy
is washed out (e.g., Acheson 1979). The stability criterion is as
given by equation (\ref{eq:acheson}), except that the thermal buoyancy
term is reduced by a factor $\eta/{\mathcal K}$. When there is no
composition gradient, the critical magnetic field $B_c$ becomes
$B_c^\prime=(\eta/{\mathcal K})^{1/2}B_c$. We plot $B_c^\prime$ as a
dotted line in Figure \ref{fig:Bcrit1}.

To estimate $\eta/{\mathcal K}$ in the ocean, we use the heat equation
(eq.~[\ref{eq:heat}]) to define a thermal time $t_{\rm
therm}=c_py^2/\rho K=3\kappa c_py^2/4acT^3$, so that $\eta/{\mathcal
K}=t_{\rm therm}/t_{\rm diff}$. The heat transport in the ocean is by
electron conduction with $K=\pi^2n_ek_B^2T\tau/3m_\star$, and we use
the results of \S \ref{sec:DIFF} for $t_{\rm diff}$, giving
\begin{equation}
{\eta\over{\mathcal K}}={2\times 10^{-4}\over T_8}\ 
{\left(1+x^2\right)^2\over x^3}\Lambda_{ei}^2\left({Z\over 30}\right).
\end{equation}
Taking $\Lambda_{ei}=1$, $g_{14}=1.9$, $\Gamma_1=4/3$ and $\mu_e=2$,
the critical magnetic field is
\begin{eqnarray}
B_c^\prime=10^9\ {\rm G}\ 
\left[{1+x^2\over g(x)^{1/2}}\right]
\left({\dot m\over 0.01\ \dot m_{\rm Edd}}\right)^{-1/2}
\left({Z\over 30}\right)^{1/2}.
\end{eqnarray}
When $x\gg 1$, we find
\begin{eqnarray}\label{eq:Bcprime}
B_c^\prime=5\times 10^9\ {\rm G}\ y_{10}^{5/8} 
\left({\dot m\over 0.01\ \dot m_{\rm Edd}}\right)^{-1/2}
\left({Z\over 30}\right)^{1/2},
\end{eqnarray}
a similar scaling with depth to $B_c$.

To summarize, we find that instability requires $B\gtrsim 10^{10}\
{\rm G}$ at the base of the ocean for $\dot m=0.01\ \dot m_{\rm Edd}$,
or $B\gtrsim 10^{11}\ {\rm G}$ for $\dot m=0.1\ \dot m_{\rm Edd}$. We
do not expect buoyancy instability to operate in the crust, as long as
the rigidity of the crust is able to counteract the buoyancy
force. The detailed outcome of the instability will depend on its
non-linear development. However, it seems reasonable to take the
critical magnetic field of equation (\ref{eq:Bcprime}) as an upper
limit on the strength of any buried field at the boundary between the
ocean and crust.

\section{Discussion and Conclusions}

We have investigated whether the magnetic field of a LMXB neutron star
may be diamagnetically screened by the accreted matter. The magnetic
profile in the outer layers is set by the competition between
downwards advection and compression by accretion, and upwards
transport by Ohmic diffusion. In steady-state, we have shown that the
magnetic field strength decreases through the outer crust and ocean by
$n$ orders of magnitude, where $n\approx \dot m/0.02\ \dot m_{\rm
Edd}$ (see \S \ref{sec:steady}). The steady-state profiles are stable
to buoyancy instabilities provided the magnetic field in the ocean and
crust is $\lesssim 10^{10}$--$10^{11}\ {\rm G}$.

\begin{figure}
\epsscale{1.0}\plotone{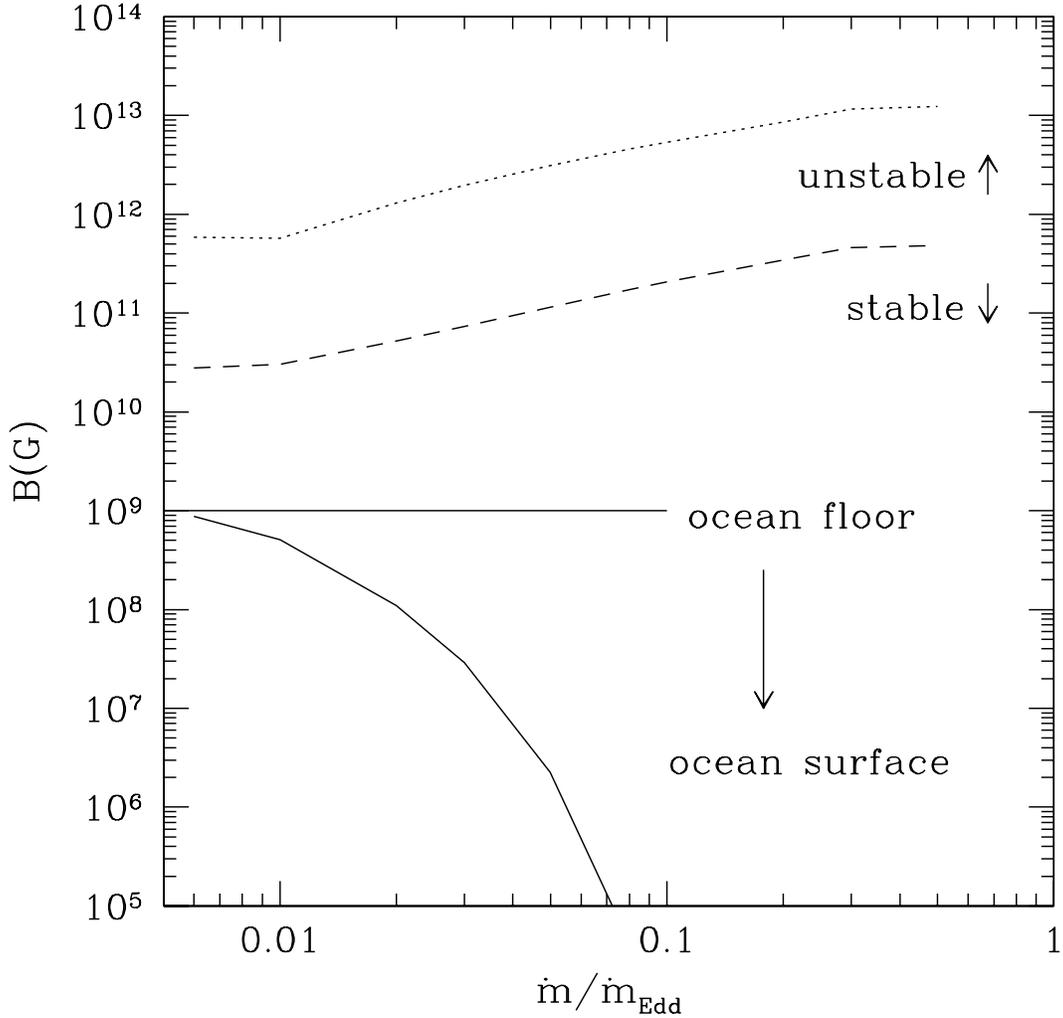}\epsscale{1.0}
\caption{A summary of our results. The solid lines indicate the
screening of the magnetic field by the ocean. Assuming a value of
$10^9\ {\rm G}$ at the base of the ocean (upper solid line), we plot
the B field at the ocean surface (lower solid line). The magnetic
field strength decreases by roughly $\dot m/0.02\ \dot m_{\rm Edd}$
orders of magnitude through the ocean. The dashed and dotted lines
show the B field needed at the ocean floor for magnetic buoyancy
instability to occur in the ocean, with and without the effects of
thermal diffusion, respectively. The dashed line gives an upper limit
to the strength of any buried field.  \label{fig:bplot}} \end{figure}

Our results are summarized in Figure \ref{fig:bplot}. Arbitrarily
choosing $B=10^9\ {\rm G}$ at the crust/ocean boundary, we plot the
magnetic field at the surface of the ocean for different accretion
rates. At large accretion rates, the magnetic field strength decreases
dramatically through the ocean, and the stellar field is screened by
the accreted layer. At low accretion rates, the magnetic field
penetrates into the accreted layer, and screening is ineffective. The
dashed and dotted lines in Figure \ref{fig:bplot} show, with and
without thermal diffusion (see \S \ref{sec:stability} for the physics
of this difference), the critical magnetic field needed at the top of
the crust for magnetic buoyancy instability to occur in the
ocean. This provides an upper limit to the strength of any buried
field, since, for larger field strengths, we expect additional upwards
transport of magnetic flux by buoyancy instabilities.

We have shown that for accretion rates $\lesssim 0.01\ \dot m_{\rm
Edd}$, microscopic Ohmic diffusion prevents screening of the neutron
star magnetic field by the accreted matter.  At larger accretion
rates, field burial may be possible; however, we stress that we have
considered only a plane-parallel model with a simple field
geometry. For example, consideration of a more complex field geometry
in a spherical shell may indicate that additional modes of magnetic
instability are available. In addition, the burial process itself is
not well-understood although some calculations of the spreading of
matter initially confined to the polar cap of a strongly magnetized
neutron star with $B\sim 10^{12}\ {\rm G}$ have been carried out
(Hameury et al.~1983; Brown \& Bildsten 1998; Litwin, Brown, \& Rosner
2001).

Our results may explain why SAX J1808.4-3658 is the only known LMXB to
show pulsations in its persistent emission (upper limits to the pulsed
fraction in other sources are $\sim 1$\%; Leahy, Elsner \& Weisskopf
1983; Mereghetti \& Grindlay 1987; Wood et al.~1991; Vaughan et
al.~1994). This source has similar X-ray spectral and variability
properties to other LMXBs (van der Klis 2000, and references therein),
but has an unusually low accretion rate. The time-averaged accretion
rate is $\approx 10^{-11}\ M_\odot\ {\rm yr^{-1}}\approx 10^{-3}\ \dot
m_{\rm Edd}$ (Chakrabarty \& Morgan 1998; Bildsten \& Chakrabarty
2001), and at the peak of outburst reaches $\lesssim 0.03\ \dot m_{\rm
Edd}$. This accretion rate is similar to that at which we expect
screening to become ineffective. This leads us to suggest that perhaps
most LMXBs have their dipole magnetic fields screened by the accreted
matter, whereas SAX J1808.4-3658 accretes slowly enough that the
magnetic field is able to rapidly penetrate the freshly-accreted
material. The other faint Galactic center transients discovered with
BeppoSAX (e.g., King 2000) may therefore be promising sources to
search for persistent pulsations.

We have concentrated on the steady-state magnetic field
profiles. However, there may be important time-dependent evolution of
the magnetic field. In particular, if the magnetic field is screened,
then we expect it to emerge once accretion ceases. The timescale for
this depends on the depth of the screening currents (Young \&
Chanmugam 1995). For the inner crust, the Ohmic diffusion time is
$10^4$--$10^5$ years, always less than the accretion time (see Figure
9 of Brown \& Bildsten 1998). Thus we expect the relevant timescale to
be the Ohmic time across the outer crust, which is extremely short,
only 100--1000 years. It therefore seems unlikely that screening alone
can explain the low magnetic fields of the millisecond radio
pulsars. Potentially observable time-dependent effects may occur in
those transiently-accreting systems whose accretion rate during
outburst is large enough for screening to be important. After an
outburst, we would expect the magnetic field profile to relax by Ohmic
diffusion on a timescale comparable to the outburst duration.

Our results show that evolution of magnetic field in the outer crust
is sensitive to the level of impurity there, as suggested by Schatz et
al.~(1999). For a primordial crust with a low level of impurity, as
assumed by previous workers, phonon scattering dominates the
conductivity. However, as shown by recent calculations of
nucleosynthesis during H/He burning in the atmosphere (Schatz et
al.~1998, 1999; Koike et al.~1999), the crust is likely very impure. In this
case, the conductivity is set by scattering from ions, leading to
shorter Ohmic dissipation times than previously assumed.

We have not discussed the layer of H/He that lies above the
ocean. Because of the different composition, the Ohmic diffusion time
in this layer is longer than that in the ocean. However, we find that
the magnetic profile in this layer is determined by a different
process, thermomagnetic drift (Urpin \& Yakovlev 1980) driven by the
heat flux from nuclear burning. The evolution of the magnetic field in
this layer is important for understanding Type I X-ray bursts. In
particular, the drifting oscillations observed during bursts with RXTE
(see Strohmayer 1999 and van der Klis 2000 for reviews) have been
interpreted as due to a differentially rotating atmosphere. However,
Cumming \& Bildsten (2000) pointed out that a poloidal field $\gtrsim
10^5\ {\rm G}$ is large enough to prevent differential rotation. Type
I X-ray bursts are thus promising probes of the magnetic field in the
surface layers. We address magnetic field evolution in the H/He layers
in a separate paper.

\acknowledgements We would like to thank Phil Arras, Steve Cowley,
Ramesh Narayan, Andreas Reissenegger, Chris Thompson, and Dmitrii
Yakovlev for useful conversations. This research was supported in part
by the National Science Foundation under Grant PHY99-07949 and
Grant AST-9800616, and by NASA
via grant NAG~5-8658. L.B. is a Cottrell Scholar of the Research
Corporation.

\end{document}